\documentclass[aps,twocolumn,reprint,prd,nofootinbib,showkeys,superscriptaddress,preprintnumbers]{revtex4-1}

\usepackage{amsmath,amssymb,graphicx,yfonts,tensor}
\usepackage{adjustbox}
\usepackage{hyperref}
\usepackage{xcolor}
\usepackage{afterpage}
\usepackage{float}
\usepackage{verbatim}
\usepackage{silence}
\graphicspath{{figures/}}

\newcommand{\pr}{\partial}
\newcommand{\be}{\begin{equation}}
\newcommand{\ee}{\end{equation}}
\newcommand{\bec}{\begin{equation*}}
\newcommand{\eec}{\end{equation*}}

\newcommand{\bea}{\begin{eqnarray}}
\newcommand{\eea}{\end{eqnarray}}

\newcommand{\om}{\omega}
\newcommand{\Om}{\Omega}

\begin{document}
\title{Electromagnetic~Antennas~for~the~Resonant~Detection of~the~Stochastic~Gravitational~Wave~Background}

\author{Nicolas Herman} \email{nicolas.herman@unamur.be}
\affiliation{ Department of Mathematics and Namur Institute for Complex Systems (naXys), University of Namur, Rue Grafé 2, B-5000, Namur, Belgium}

\author{L\'eonard Lehoucq}
\email{lehoucq@iap.fr}
\affiliation{Institut d’Astrophysique de Paris, Sorbonne Université and CNRS, UMR 7095, 98 bis bd Arago, F-75014 Paris, France}

\author{Andr\'e F\H{u}zfa} \email{andre.fuzfa@unamur.be}
\affiliation{ Department of Mathematics and Namur Institute for Complex Systems (naXys), University of Namur, Rue Grafé 2, B-5000, Namur, Belgium}

%\author{Sebastien Clesse}  \email{sebastien.clesse@ulb.ac.be}
%\affiliation{Service de Physique Th\'eorique, Universit\'e Libre de Bruxelles (ULB), Boulevard du Triomphe, CP225, B-1050 Brussels, Belgium}

\date{\today}

\begin{abstract}
Some stochastic gravitational wave background models from the early Universe has a cut-off frequency close to 100 MHz, due to the horizon of the inflationary phase. To detect gravitational waves at such frequencies, resonant electromagnetic cavities are very suitable. In this work, we study the expected frequency response of such detectors using a brand new approach, and show how we could use them to probe this cut-off frequency and also the energy density per frequency of this stochastic background. This paper paves the way for further experimental studies to probe the most ancient relic of the Universe. 
\end{abstract}

\pacs{}
\maketitle
\section*{Introduction}
The stochastic gravitational wave background (SGWB) is the most ancient relic of the Big Bang. This is the analog of Cosmic Microwave Background (CMB) for gravitational waves (GWs), potentially giving information about the early ages of the Universe, before the formation of atoms and nuclei, glimpsing directly when fundamental interactions supposedly splitted. Maggiore~\cite{Maggiore2007} described SGWB first as isotropic, gaussian and stationary, and its frequency dependence is contained in a one-sided power spectral density. Recent works have detailed SGWB at high frequencies~\cite{Ringwald2021,Ghiglieri2015,review,Sousa2020,Ringwald:2022xif}. In the review~\cite{review}, a variety of possible sources of stochastic background are considered, which mostly arises from hypothetical physics of the early universe : (pre)heating, oscillons, cosmic strings, inflation, to name but a few. Most of these hypothetical sources can be characterized by two parameters, first the energy density $\Om_{GW}$ per logarithmic frequency sampling, and then some of potential sources have a cut-off frequency in the MHz-GHz band. This is due to GWs trapped in the horizon at the epoch of the end of inflation~\cite{Maggiore2}. To detect GWs at such high frequency, Electromagnetic (EM) detectors should be considered. Their working principle is based on wave resonance mechanism and was discovered by Gertsenshtein~\cite{gertsenshtein}, although this author worked it out for GW generation. Detection can be achieved with the so-called inverse Gertsenshtein effect, which can be physically described as follows. A GW fundamentally constitutes of a local volume distortion . If we put a magnetic field on the way of this spacetime distortion, the passing GW will modify the EM flux by affecting the volume, giving rise to an induced EM field from Lenz's principle. This induced EM field betrays the passage of a GW, as it inherits the frequency from its gravitational progenitor, all features that constitute a specific response of EM detectors based on inverse Gertsenshtein effect.
Some detector proposals were made just few years after Gersenshtein's discovery~\cite{boccaletti,braginskii1,braginskii2,grishchuk1,grishchuk2,grishchuk2bis,delogi,pegoraro1,pegoraro2,caves,cruise}. Recently, this topic of EM detection of GW has seen a renewal of interest after first GW detection by LIGO~\cite{LIGO}, with proposals based on the inverse Gertsenshtein mechanism~\cite{Zheng2018,Ejlli2019,Berlin2021,Domcke2021,Domcke2022}.
These kind of detectors are complementary with other detection techniques to detect the whole spectrum of GWs. For instance, interferometers can detect GWs in the mHz to kHz band. EM detectors can theoretically detect any GW frequency, but the induced field intensity and detector dimensions make it suitable for high frequencies, from kHZ to THz. 
Research on (Ultra-)High Frequency GW is currently active. The review~\cite{review} lists potential sources and detectors for those frequencies. 
We focus here particularly on resonant EM detectors, already described in~\cite{Herman2021}, by treating directly the detectors frequency response and apply this to the experimental SGWB search.
Resonant detectors are indeed suitable to spot on a narrow frequency range, and are especially interesting to isolate the cut-off frequency of SGWB. 
We claim resonant EM detectors of GWs are promising tools for the detection of SGWB, allowing to reveal the position of its power spectrum cut-off frequency and the variation of the related cosmic energy density with the frequency.

\section{Proposed experimental setups} 
SGWB detection antennas are based on the conversion of GWs into EM fields through the inverse Gertsenshtein effect. Precisely, we consider resonant detection schemes where the interaction of the passing GWs with some external magnetic field induces excitation of EM modes into a cavity. The energy induced into the resonator by the passing GW is faint, due to the weakness of the gravitational coupling. However, the root-mean square (rms) value of the induced power inside the cavity is proportional to the strain of the incoming GW.
More details about detection scheme configuration and observable can be found
in patents~\cite{patent} or in article~\cite{Herman2021}. 
The fundamental equation that rules the conversion of GWs to EM field is the Maxwell wave equation on curved spacetime, from which one can derive the induced EM fields.
Let us consider the detection schemes have cylindrical symmetry. The electromagnetic field within the resonant cavity is a superposition of some background static magnetic field $\vec{B}^{(0)}$ and a perturbation which is the induced electromagnetic field due to the Gertsenshtein effect $\vec{B}^{(1)}$. %We shall also assume that some ohmic losses happen at cavity walls (see also ~\cite{grishchuk2bis}), so there is a current density $\vec{j}_{\rm loss}=\frac{\omega}{\epsilon_0 Q}\vec{E}^{(1)}$ where $\vec{E}^{(1)}$ is the induced electric field, $\omega$ its frequency. In the above relation, $Q$ is the quality factor of the cavity which we assume to be around $10^5$, an order of magnitude that is consistent with resonant cavities currently used for axion\cite{ADMX2}.
 Under these assumptions, one obtains the following wave equation for the induced magnetic field $\vec{B}^{(1)}$,
\be
\left(-\frac{1}{c^2}\frac{\partial^2}{\pr t^2}+\vec{\Delta}\right) \vec{B}^{(1)}=-\mu_0\vec{\nabla} \times \left[\vec{\mathcal{J}}_{\rm eff}(\vec{B}^{(0)},h_{\mu\nu})+\vec{\mathcal{J}}_{\rm loss}\right] 
\label{eq:wavej}
\ee
where $\vec{\Delta}$ is the Laplacian and
$\vec{\mathcal{J}}_{\rm eff}(\vec{B}^{(0)},h_{\mu\nu})$ is an effective current density that source the induced magnetic field, containing the correction due to local modification of volume caused by the passing GW and $\vec{\mathcal{J}}_{\rm loss}$ denotes the ohmic losses in the cavity. We will discuss and describe in the next section the computation of the effective current density. The boundary condition for the magnetic field is $\vec{B}^{(1)}_\perp=0.$ An analogous equation of Eq.~(\ref{eq:wavej}) can be written for the induced electric field $\vec{E}^{(1)}$. However, the leading effect is given by the induced magnetic perturbation which is amplified by the external magnetic field. Indeed,
 the variation of energy $\Delta\mathcal{E}$  inside the cavity of volume $V$ at first order in the induced magnetic field $\vec{B}^{(1)}$
 is given by
\be 
\Delta \mathcal{E} \approx \frac{1}{\mu_0}\int_V \vec{B}^{(0)} \cdot \vec{B}^{(1)} dV,
\label{eq:dE}
\ee 
which is therefore boosted by the external magnetic field $\vec{B}^{(0)}$. This energy variation evolves with time with a similar frequency content as the passing GW and its rms amplitude is directly proportional to the GW strain (see \cite{Herman2021}).
Consequently, we focus here on the induced magnetic field, which constitutes the dominant effect in the induced rms power in the cavity, obtained by timely derivating the energy fluctuation Eq.~(\ref{eq:dE}). This variation at first order is very promising for high-frequency GW detection since the order of magnitude of the induced EM power is much higher than other proposals~\cite{Ejlli2019,Berlin2021,Zheng2018}. 
% AF
These proposals focused on second order effects in the induced electromagnetic fields (and therefore in GW too). The work~\cite{Berlin2021} even discards first order effects on the basis of their vanishing time average.
However, it must be reminded that time average is not physically relevant for energy transport: periodic signals do carry energy by non-vanishing rms average. This simple physical fact lies at the very basis of the alternative current (AC) technology that we are using everyday. Fundamentally, the carried energy is related to the $L^2-$norm of time-dependent signals (the rms average), given by the power spectrum in Fourier analysis (since such transform is unitary).
Therefore, the quadratic mean of the incoming GW can also be measured in the energy fluctuations Eq.~(\ref{eq:dE}) and, since these are amplified by the strong external magnetic field, they constitute a dominant effect compared to other interesting proposals ~\cite{Ejlli2019,Berlin2021,Zheng2018}. By analyzing the Eq.~(\ref{eq:dE}), we can see that we only need to consider the induced field along the external magnetic field direction, which is here assumed transverse to the longitudinal axis of the resonant cavity. Therefore, we will compute the source term $\vec{\mathcal{J}}_{\rm eff}$ in this way.

The wave equation~(\ref{eq:wavej}) can be decomposed in eigenmodes, as in~\cite{grishchuk2bis}, by developing the solution on the eigenfunctions of the Laplacian operator with required boundary conditions on the cavity walls. These functions are{\footnotesize \begin{eqnarray}
	\psi^r_{kmn}&=&C^r_{kmn}\cdot\frac{\mathcal{R}_{km}(r)}{mr}\cdot 
	\begin{Bmatrix}
		\cos\\
		\sin
	\end{Bmatrix}
	\left(m\theta\right)
	\cdot 
	\begin{Bmatrix}
		\cos\\
		\sin
	\end{Bmatrix}
	\left(\frac{2\pi n z}{L}\right)\label{eq:psir}\\
	\psi^\theta_{kmn}&=& C^\theta_{kmn}\cdot\frac{d\mathcal{R}_{km}(r)}{dr}\cdot 
	\begin{Bmatrix}
		-\sin\\
		\cos
	\end{Bmatrix}
	\left(m\theta\right)
	\cdot 
	\begin{Bmatrix}
		\cos\\
		\sin
	\end{Bmatrix}
	\left(\frac{2\pi n z}{L}\right)\label{eq:psitheta}\\
	\psi^z_{kmn}&=&C^z_{kmn}\cdot\mathcal{R}_{km}(r)\cdot 
	\begin{Bmatrix}
		\cos\\
		\sin
	\end{Bmatrix}
	\left(m\theta\right)
	\cdot 
	\begin{Bmatrix}
		\sin\\
		\cos
	\end{Bmatrix}\left(\frac{2\pi n z}{L}\right)\label{eq:psiz},
\end{eqnarray}

}
where $k,m,n$ are integers that appears during the variable separation method. These are "quantum" numbers to differentiate each harmonic. $C_{kmn}$ are normalization constants and $L$ is the length of the cavity. The radial function $\mathcal{R}_{km}(r)$ depends if we consider the TM or te TEM cavity, and contains a combination of Bessel functions that respect the boundary conditions. The roots of these functions are denoted by $\alpha_k$ and can be tuned by the cavity geometry (see \cite{Herman2021}). These eigenfunctions satisfy the Helmholtz equation
\begin{equation}
	\vec{\Delta} \vec{\psi}_{kmn}=-\Omega^2_{kmn}\vec{\psi}_{kmn},
	\label{eq:helmholtz}  
\end{equation}

where
\begin{equation}
	\Omega^2_{kmn}=\alpha^2_{k}+\frac{4\pi^2 n^2}{L^2}.
	\label{eq:omegakmn}
\end{equation}
The wavenumbers $\Omega_{kmn}$ are the resonant wavenumbers of the cavity. We can obtain the resonant frequencies by dividing the wavenumbers by $c$.
% \subsection{Calculating the RMS induced power of the detectors with a frequency approach}
With the spectral decomposition \begin{eqnarray}
	{B^{(1) r,\theta,z}}(t,\vec{r})&=&\sum_{k,m,n} \hat{b}_{kmn}^{r,\theta,z} (t) \psi_{kmn}^{r,\theta,z}(\vec{r})\label{eq:Bdecomp},\\
	\mu_0\left(\vec{\nabla}\times\vec{j^{\rm eff}}\right)^{r,\theta,z}(t,\vec{r})&=&\sum_{k,m,n} \hat{s}_{kmn}^{r,\theta,z} (t)  \psi_{kmn}^{r,\theta,z}(\vec{r}),
\end{eqnarray} the inhomogeneous wave equation~(\ref{eq:wavej}) becomes a forced and damped harmonic oscillator equation for each eigenmode $\hat{b}_{kmn}^{r,\theta,z}$. The damping term comes from the ohmic losses that we can express as $\vec{\mathcal{J}}_{\rm loss}=\sigma \vec{E}^{(1)}$ and using the Maxwell equation $\pr_t\vec{B}^{(1)} =  -\vec\nabla\times\vec{E}^{(1)}$ at first order. These oscillator equations are 
\begin{equation}
	\frac{1}{c^2} \frac{d^2  \hat{b}_{kmn}^{r,\theta,z}}{dt^2} +\frac{\Omega_{kmn}}{c Q}\frac{d  \hat{b}_{kmn}^{r,\theta,z}}{dt}+ \Omega^2_{kmn}  \hat{b}_{kmn}^{r,\theta,z} =   \hat{s}_{kmn}^{r,\theta,z} 
	\label{eq:harmosc}
\end{equation}
for each $(k,m,n)$ mode, where we express the ohmic losses effective conductivity $\sigma$ for each mode as $\sigma=\frac{c\Omega_{kmn} \epsilon_0}{Q}$.
Only purely radial modes survives to the volume integral giving the induced energy Eq.~(\ref{eq:dE}): those which are not constant along the longitudinal axis and not behaving as a sinusoidal function of the azimuth angle will disappear due to cylindrical symmetry (shown in \cite{Herman2021}). In other words, only the radial modes $\hat{b}_{k10}^{r,\theta,z}(t)$ contributes to the total energy variation. We are going to use the temporal Fourier transform, that we can apply to any time-dependent function $f(t)$, and obtain the temporal Fourier transform \begin{equation}
	\tilde{f}(\omega)=\int_{-\infty}^{\infty} f(t) e^{-i\omega t} dt,
\end{equation} that depends on $\omega=2\pi\nu$ where $\nu$ is the frequency. This temporal Fourier transform will turn our harmonic oscillator differential equation in an algebraic one and will allow us to compute the frequency response of the cavity. Thus our harmonic oscillator equation (\ref{eq:harmosc}) becomes
\begin{equation}
	-\frac{\omega^2}{c^2} \tilde{b}_{kmn}^{r,\theta,z} + \frac{i\omega \Omega_{kmn}}{cQ} \tilde{b}_{kmn}^{r,\theta,z} +\Omega^2_{kmn}  \tilde{b}_{kmn}^{r,\theta,z}=\tilde{s}_{kmn}^{r,\theta,z} \label{eq:harmoscf}
\end{equation}
where $\tilde{b}_{kmn}^{r,\theta,z} $ is the temporal Fourier transform of $\hat{b}_{kmn}^{r,\theta,z}$, and 
\begin{equation}
	\tilde{s}_{kmn}^{r,\theta,z}(\omega)= \int_V \mu_0\left(\vec{\nabla}\times\vec{{J}}^{\rm eff}(\omega,\vec{r})\right)^{r,\theta,z} \psi_{kmn}^{r,\theta,z}(\vec{r}) dV, \label{eq:skmnf}
\end{equation}
where $\vec{{J}}^{\rm eff}(\omega,\vec{r})$ is the temporal Fourier transform of $\vec{\mathcal{J}^{\rm eff}}(t,\vec{r})$. This can be made because the temporal and the spatial variables in our effective current density are independent. The solution for each mode is given by
\begin{equation}
	%\hat{b}_{k10} (t)=e^{-\frac{\alpha_k c}{2Q} t} \left(A_{k10}  \cos\left(\varpi_k t\right) + B_{k10}  \sin\left(\varpi_k t\right)\right) + \frac{1}{2\pi}\int_{-\infty}^{\infty}  \frac{\tilde{s}_{k10} (\omega)}{\alpha_k^2+\frac{i\omega\alpha_k c}{Q}-\omega^2} e^{i\omega t} d\omega
	\tilde{b}_{k10}^{r,\theta,z} (\omega)=\frac{A_{k10}^{r,\theta,z} +i B_{k10}^{r,\theta,z}  }{-\frac{\alpha_k c}{2Q}+i\left(\omega-\varpi_k\right)}+ \frac{c^2\tilde{s}_{k10}^{r,\theta,z} (\omega)}{c^2\alpha_k^2+\frac{i\omega\alpha_k c}{Q}-\omega^2}
	\label{eq:solk10f}
\end{equation}
with $\varpi_k=c\alpha_k\sqrt{1-\frac{1}{2Q}}$ for any $\omega>0$. The constants $A_{k10}^{r,\theta,z} $ and $B_{k10}^{r,\theta,z} $ are determined with initial condition $\hat{b}_{k10}^{r,\theta,z}(t=0)=\left.\frac{d\hat{b}_{k10}^{r,\theta,z}}{dt}\right\vert_{t=0}=0$
Using the previous equation and assuming the external field aligned with the X-axis, $\vec{B^{(0)}}=B_0 \vec{e_x}$ we can finally compute the induced power in the cavity at first order of the strain, depending on the frequency $\omega$,
\begin{equation}
	\mathcal{\tilde{P}}(\omega)=\frac{iB_{0}\omega}{\mu_0}\int_V  \tilde{B}^{(1)}_x dV,
	\label{eq:Pinf}
\end{equation}
where $\tilde{B}^{(1)}_x$ is the temporal Fourier transform of $B^{(1)}_x$. Passing the expression of $\tilde{B}^{(1)}_x$ in cylindrical coordinates, and using the cylindrical harmonics decomposition (\ref{eq:Bdecomp}), we obtain that the induced power in our cavity is
\be
\mathcal{\tilde{P}}(\omega)=\frac{2i \pi \omega B_0}{\mu_0}\sum_{k}I_k \tilde{b}_{k10}(\omega),
\label{eq:P(w)}
\ee
where $ \tilde{b}_{k10}(\omega)=\tilde{b}_{k10}^r(\omega)=-\tilde{b}_{k10}^\theta(\omega)$ and the radial integral $I_k$ is given by 
\be
I_k =\int_{0,R_1}^{R,R_2} R_{k1}(r) dr.
\ee
The bounds of the integral depends if we consider a TM or a TEM cavity. The equation (\ref{eq:P(w)}) give us a very powerful way to compute the induced power in our cavity at first order, by only using the harmonic oscillator solution (\ref{eq:solk10f}). The only remaining development we need to inverstigate is expressing the forcing term given by the equation (\ref{eq:skmnf}), that will be done in the following.

\section{Computing the source term and the RMS impulse response}
Before providing an analysis of the frequency response of such detection schemes, we have to discuss the choice of the frame modeling the detection process.  A simplification made in the article~\cite{Herman2021}, as emphasized in subsequent work ~\cite{Berlin2021}, was that we consider the traceless-transverse frame (TT gauge) for the incoming GW and the proper detector frame for the electromagnetic fields involved in the process. In the present work, we first investigate the validity of this approximation by computing explicitly the gauge transformation from traceless-transverse coordinates to the Fermi-Normal ones. We then give both analytical and numerical evidence in favor of the simplification used in the work  ~\cite{Herman2021}.
%deals with this mix of frames and we will model the whole detection process in the detector frame.
%will deduce the  induced RMS power of the cavity at first order in the GW amplitude. 
The passing GW coming from astrophysical sources is usually described in traceless-transverse gauge, where $h^{\rm TT}_{+,\times}$ denotes the polarization of the propagating transverse modes. However, this coordinate choice, although suitable for GW propagation in vacuum is not ideal for describing the detection scheme. For the last, a coordinate choice based on the electromagnetic fields at play, the proper detector frame, is more convenient.
One has therefore to express the incoming GW (given in the TT gauge) in the proper detector frame, as noticed in~\cite{Berlin2021}. This new frame can be described physically as the coordinates where an inertial observer will measure physical quantites in his vicinity. %, although this was not fully explicited in this work.
 Let us carefully perform this transformation here by following the results obtained in Refs.~\cite{Marzlin1994,Rakhmanov_2014}.  Particularly, we can use the development of the metric perturbation in the Fermi-Normal coordinates made by Rakhmanov~\cite{Rakhmanov_2014}.
 %These coordinates form a unique and continuous set of non-rotating coordinates, where the spacetime is flat. The tetrad that form the base of this coordinate system is parallel transpoted along the geodesics for an inertial observer. 
 In this coordinate system, where a gravitational plane wave propagating along the z-axis, the $h_{\mu \nu}$ tensor can be expressed only with the spatial coordinates and the following functions
\bea
P_{+,\times}(z,t)&=&\sum_{n=2}^\infty \frac{n-1}{(n+1)!}z^n \frac{d^n h^{\rm TT}_{+,\times}(t)}{dt^n} \label{eq:P}\\
Q_{+,\times}(z,t)&=&\sum_{n=2}^\infty \frac{n}{(n+1)!}z^n \frac{d^n h^{\rm TT}_{+,\times}(t)}{dt^n} \label{eq:Q}
\eea
where $z$ is the coordinate related to the propagation direction of the GW in the detector frame and $t$ is the time in the detector frame. 

The metric perturbation computed by Rakhmanov~\cite{Rakhmanov_2014} has the form
\begin{eqnarray}
	h_{11}&=&P_+, \label{eq:first}\\
	h_{22}&=&-P_+,\\
	h_{12}&=&P_\times,\\
	h_{13}&=&-\frac{1}{z}\left(xP_+ + yP_\times\right),\\
	h_{23}&=&-\frac{1}{z}\left(xP_\times - yP_+\right),\\
	h_{33}&=&\frac{1}{z^2} \left(\left[x^2-y^2\right]P_+ +2xy P_\times\right),\\
	h_{01}&=&-\frac{1}{z}\left(xQ_+ + yQ_\times\right),\\
	h_{02}&=&-\frac{1}{z}\left(xQ_\times - yQ_+\right),\\
	h_{03}&=&\frac{1}{z^2} \left(\left[x^2-y^2\right]Q_+ +2xy Q_\times\right),\\
	h_{00}&=& 2 h_{03} - h_{33} \label{eq:last},
\end{eqnarray}

% AF:
%Ici il faut remanier. Tu dois expliquer le principe du changement de frames. Si je comprends bien, les deux frames coïncident en un point, dans le référentiel localement inertiel puis se différencient lorsque tu t'éloignes de l'origine où on a placé l'observateur. Dans la TT gauge, on a deux polarisations qui vont donc sourcer les 10 perturbations du detector frame. On obtient ces perturbations du detector frame par développement en série construit à partir des fonctions P et Q. Ce sont ces perturbations de la métrique dans ce detector frame qui vont être la source de l'effet Gertsenshtein inverse. Tu peux alors l'introduire après cette explication. 
% AF
 Since we work in the proper detector frame, 
%the source term explained in~\cite{Herman2021} is no longer valid : ne te devalorise pas, tu finis par montrer que ça ne change pas grand chose!
 the metric perturbation tensor is generally not traceless and is not ruled by Lorenz gauge condition. As in~\cite{Berlin2021}, the effective current density is given by
\be
\begin{aligned}
\mathcal{J}_{\rm eff}^\mu&=\pr_\nu h^\mu_\alpha F^\nu_\alpha + \pr_\nu h^\nu_\alpha F^{\mu\alpha} + \frac{1}{2} \pr_\nu h F^{\mu \nu}\\
&=j_1^\mu+j_2^\mu+j_3^\mu
\end{aligned}
\ee
where $F^{\mu \nu}$ is the Faraday tensor of the background electromagnetic field (not to be confused with the one induced by inverse Gertsenshtein effect). The general source term obtained in the detector frame is more complicated than the one in~\cite{Herman2021}, because the metric perturbations do not verify the TT gauge constraints. However,
let us show a practical analytical approximation. First, let us assume that our source term $\mu_0\left(\vec{\nabla}\times\vec{{J}}^{\rm eff}(\omega,\vec{r})\right)$ can only depend on $(z,\omega)$, as a plane wave approximation. We can show that in our case the forcing term of our oscillator equations, due to vanishing quantities in the volume integral and the source term expression in cylindrical coordinates, we can rewrite equation (\ref{eq:skmnf}) as
\be
\begin{aligned}
	\tilde{s}_{k10}(\omega)&=\tilde{s}_{k10}^r(\omega)=-\tilde{s}_{k10}^\theta(\omega)\\&= \pi I_k \int_{-\frac{L}{2}}^{\frac{L}{2}} \mu_0\left(\vec{\nabla}\times\vec{{J}}^{\rm eff}(\omega,z)\right)^{x} \, dz,
\end{aligned} \label{eq:skmnsumf}
\ee
 The metric perturbations in the proper detector frame can be obtained by equations (\ref{eq:first}) to (\ref{eq:last}) from~\cite{Rakhmanov_2014}. If we truncate the series development of the functions $P_{+,\times}$ and $Q_{+,\times}$ at the dominant $n=2$ term, one finds that
\begin{eqnarray}
	\mu_0\left(\vec{\nabla}\times\vec{j_1}\right)^{x}&=& -B_0 \frac{1}{c^2} \frac{d^2h_{+}}{dt^2} ,\\
	\mu_0\left(\vec{\nabla}\times\vec{j_2}\right)^{x}&=& \frac{B_0}{6} \frac{1}{c^2} \frac{d^2h_{+}}{dt^2} - \frac{B_0z}{3} \frac{1}{c^3} \frac{d^3h_{+}}{dt^3} , \\
	\mu_0\left(\vec{\nabla}\times\vec{j_3}\right)^{x}&=& -\frac{B_0}{3} \frac{1}{c^2} \frac{d^2h_{+}}{dt^2}.
\end{eqnarray}
If we neglect the term in $\frac{1}{c^3}$ we find out that
\begin{equation}
	\mu_0\left(\vec{\nabla}\times\vec{\mathcal{J}_{\rm eff}}\right)^{x} = -\frac{7B_0}{6} \frac{1}{c^2} \frac{d^2h_{+}}{dt^2} = \frac{7}{6} \mu_0\left(\vec{\nabla}\times\vec{j_1}\right)^{x}.
\end{equation}

That means that we can approximate the source term by using the definition of $\vec{j_1}$,
\bea
	\mu_0\left(\vec{\nabla}\times\vec{\mathcal{J}_{\rm eff}}\right)^{x} &\approx&  \frac{7}{6} \mu_0\left(\vec{\nabla}\times\vec{j_1}\right)^{x}\label{eq:approx}
\eea
Hence, we find that the source term $\vec{j_1}$, used in ~\cite{Herman2021}, only underestimates the total effective current density by $14\%$ at the leading order in the transformation from the 
traceless-transverse coordinates to the detector frame. One can now asks how far 
this analytical approximation holds valid at higher orders in the series expansion. We can therefore extend the gauge transformation to all orders in the series of Eqs.~(\ref{eq:P},\ref{eq:Q}) 
and examine numerically the convergence. Fig.~\ref{fig:my_label} in appendix gives a comparison of the effective current densities due to the metric perturbations in the proper detector frame obtained with the series expansion of ~\cite{Rakhmanov_2014}, truncated at tenth order ($n=10$). Please note that this approximation was made under the assumption that the GW propagation axis is aligned with the symmetry axis of the cavity. The directionality analysis is still to perform for further experimental development. Anyway we can see this approximation as an upper limit since the cavity response should be maximal in this configuration. Moreover, to cope with this directionality problem, one can set up two sepereate cavities with their symmetry axis perpendicular, in order to maximize the potential astrophysical GW detection. \\

With the approximation Eq.~(\ref{eq:approx}), we can use as the source term for the x-component of Eq.~(\ref{eq:wavej}) the following expression
\be
\mu_0\left(\vec{\nabla} \times \vec{\mathcal{J}}^{\rm eff}\right)^x= -\frac{7B_0}{6} \left(\frac{\pr^2 P_+}{\pr z^2}+\frac{2}{z}\frac{\pr P_+}{\pr z}\right), 
\label{eq:approx2}
\ee
that we can use for all orders in the series expansion (\ref{eq:P}).
Since the approximation~(\ref{eq:approx2}) is valid, we can use the equations~(\ref{eq:skmnsumf}) and (\ref{eq:P}) to get the source terms our our harmonic oscillator equations, we obtain then
\begin{equation}
	\begin{aligned}
	\tilde{s}_{k10}(\omega)&= -\frac{7\pi B_0 I_k}{6} \\ &\sum_{n=2}^{+\infty} \frac{1}{\left(n-1\right)\text{!} \, \,} \left[\left(\frac{L}{2}\right)^{n-1}-\left(-\frac{L}{2}\right)^{n-1}\right] \left(\frac{i\omega}{c}\right)^n \tilde{h}_+.
		\end{aligned}
\end{equation}
where $\Tilde{h}_+(\om)$ is the Fourier transform of the $+$ polarization of the incoming GW, assumed as a plane wave (see also \cite{Herman2021}).
This equation above, after small algebraic manipulations and sum indices modifications,
 we can obtain as source term
\begin{equation}
	\tilde{s}_{k10}(\omega)= \frac{7\pi B_0 I_k\omega \tilde{h}_+ }{3c} \sin\left(\frac{\omega L}{2c}\right).
	\label{eq:sk10f}
\end{equation}

%AF: est-ce que cette equation est utile pour comprendre la suite? Pour le calcul de la reponse frequentielle peut-etre?
Please note that this equation is exactly $\frac{7}{6}$ of the temporal Fourier transform of the harmonic oscillator source term computed in~\cite{Herman2021}. 

\begin{figure}
    \centering
    \includegraphics[scale=0.37]{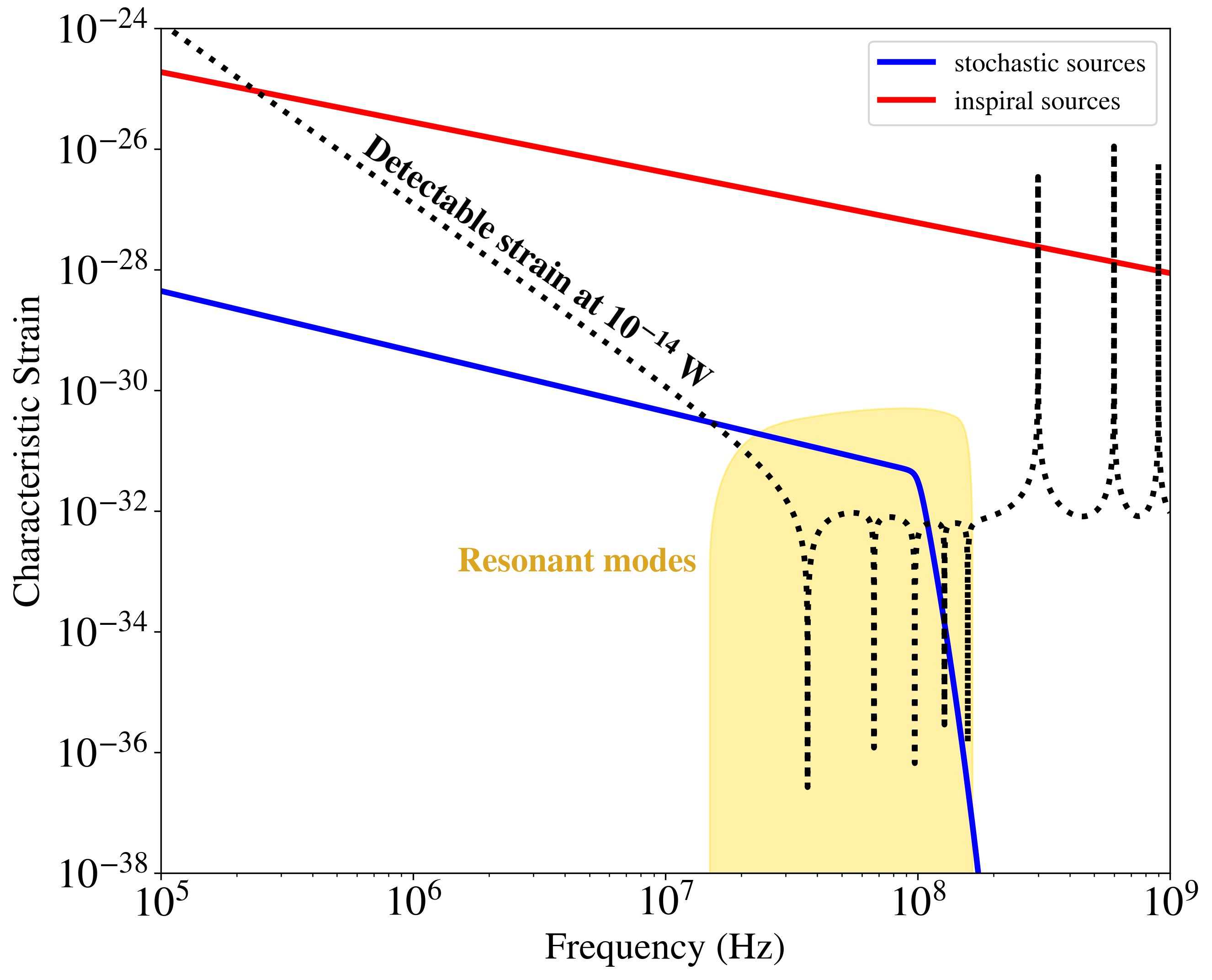}
    \caption{Expected strain sensitivity of our cavity and possible sources. The black dashed curve is the strain sensitivity for a monochromatic GW if we could detect a rms induced power of $10^{-14}$ W. We can clearly see that the detection will be better around the resonant frequencies of the cavity. The behavior at high frequency is due to the sinusoidal term in Eq.~(\ref{eq:RIR}). The blue and red lines are respectively the characteristic strain for a stochastic GW signal and an inspiral GW signal at ISCO (Innermost Stable Circular Orbit) frequency. Both curves are consistent with~\cite{review}.}
    \label{fig:detlim}
\end{figure}

 Using these computations and some results of~\cite{Herman2021} we can easily compute the expression of the root mean-square (rms) power for a monochromatic unit strain amplitude gravitational wave, the rms impulse response (RIR) of our cavity. We detailed these computation in the appendix. This rms impulse response has the form
\begin{equation}
	P_{\rm RIR}(\omega)=\frac{7\sqrt{2}c}{3\mu_0}\pi^2 B_0^2 \omega^2\left\vert\sin\left(\frac{\omega L}{2c}\right)\right\vert \mathcal{S},
	\label{eq:RIR}
\end{equation}
where
\be
\mathcal{S} =\sum_{k} \frac{\left( I_k\right)^2}{\sqrt{\left(\alpha_k^2c^2-\omega^2\right)^2+\left(\frac{\alpha_k \omega c}{2Q}\right)^2}}\cdot
\label{eq:S}
\ee
We can see the similarity with the results obtained in~\cite{Herman2021} with a direct temporal domain approach, beside the frame discussion mentioned above. We can also easily compute the rms induced power for inspiral signals, such as merging planetary primordial black holes. For an incoming GW with Fourier transform $\Tilde{h}_+(\om)$, the rms induced power can be expressed by
\begin{equation}
	\mathcal{P}_{\rm RMS}^2\simeq \int_0^\infty \vert P_{\rm RIR}(\omega) \vert^2 \vert \tilde{h}_+(\omega) \vert^2 d\omega.
	\label{eq:RMSestim}
\end{equation}
We can also obtain the expected detection scheme strain sensitivity from this impulse response diagram in Fig.~\ref{fig:detlim} where the dashed black curve shows the strain we could possibly detect with the proposed detection scheme, assuming that we could detect an induced rms power of $10^{-14} W$. The parameters chosen for our detection scheme are a $5$T external magnetic field, for a one-meter long cavity with a 5m radius. We can see a better sensitivity at the cavity resonant frequencies. Their locations depend on the radius of the cavity~\cite{Herman2021}. We also add to this figure the characteristic strain of inspiral GW signals and the one from the stochastic GW background we consider.  Both types of sources can hopefully be detected by such resonant detection schemes. 
To go further on the detection schemes modeling, the exact experimental design and its noise sources should be studied and considered. Another important point is the measurement of the rms induced power at first order. Experience gained in the field of haloscopes, like in the Axion Dark Matter Experiment \cite{ADMX}, are of direct application for the EM detection of SGWB and in general high-frequency gravitational waves. In the ADMX experiment, the rms power they can measure is of the order of $10^{-21} W$, but the time scale is quite different in our case. Their scale is order of a year, ours is microseconds for inspiral signals, and could be longer for the stochastic ones.

\section{Results on a Stochastic GW Background toy model} 
\begin{figure}
    \centering
    \includegraphics[scale=0.37]{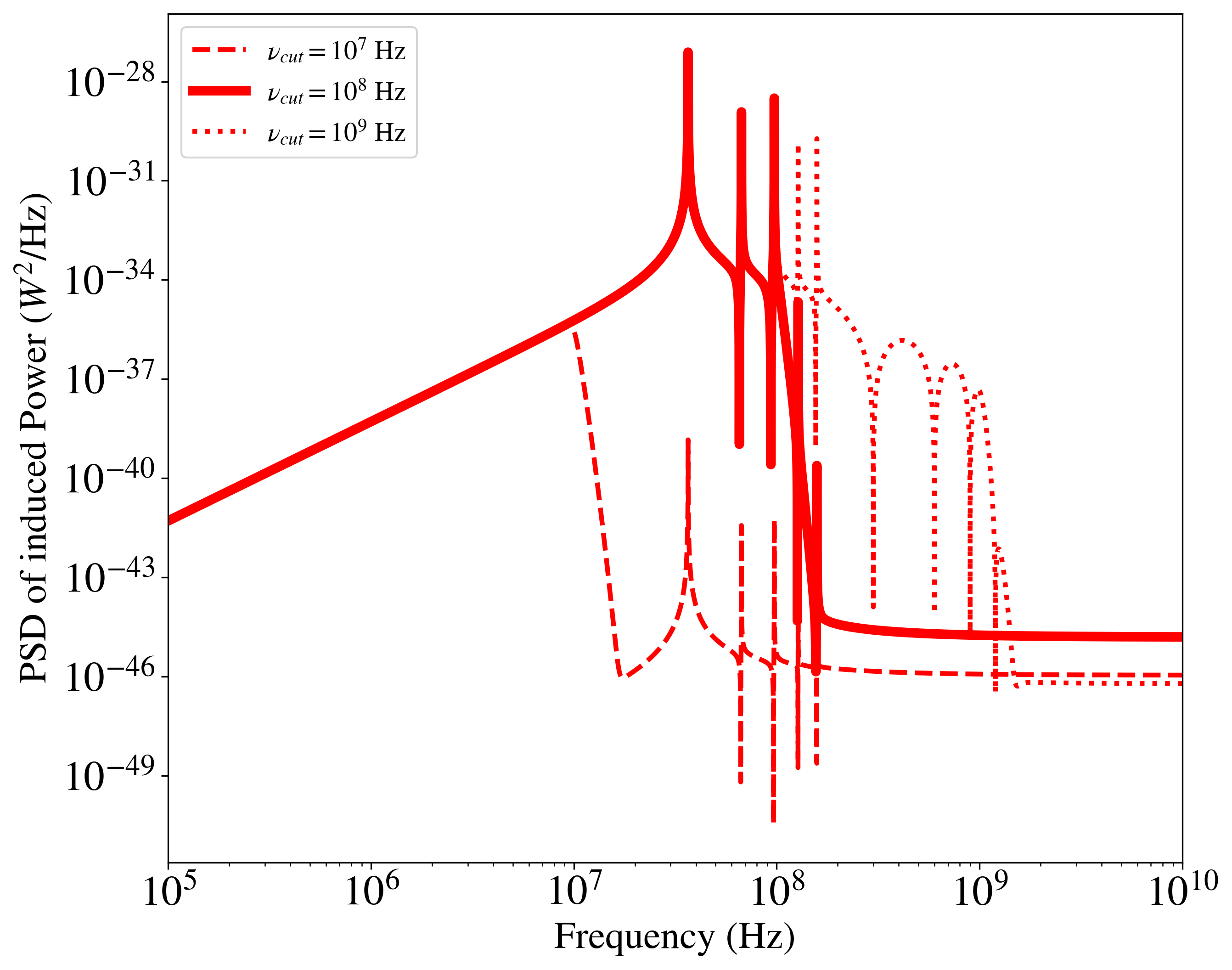}
    \caption{The power spectral density of the induced EM Power that detects our stochastic GW background toy model, for three different cut-off frequencies. We can see that the response is several orders of magnitude lower above the cut-off frequency. The cut-off frequency should appear clearly when several signals are correlated, with resonant frequencies chosen carefully.}
    \label{fig:psd}
\end{figure}

\begin{figure}
    \centering
    \includegraphics[scale=0.37]{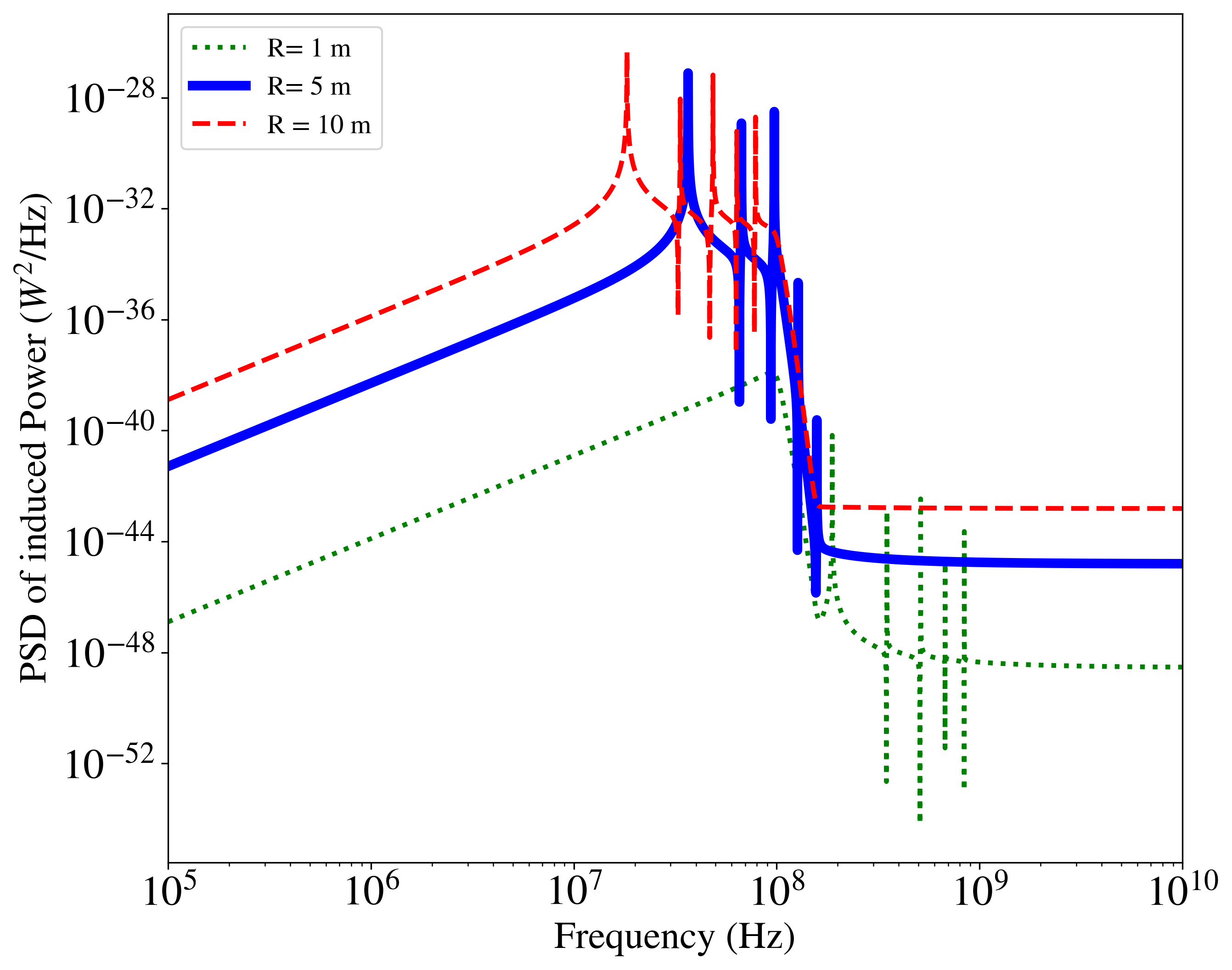}
    \caption{The power spectral density of the induced EM Power that detects our stochastic GW background toy model, for three cavity radii. The cavity resonant frequencies are depending on the radius of the cavity as shown in~\cite{Herman2021}. We can see that the cut-off frequency appear clearly when the resonant frequencies are chosen carefully. Otherwise, locate the cut-off frequency should be more difficult.}
    \label{fig:psdr}
\end{figure}

\begin{figure}
    \centering
    \includegraphics[scale=0.37]{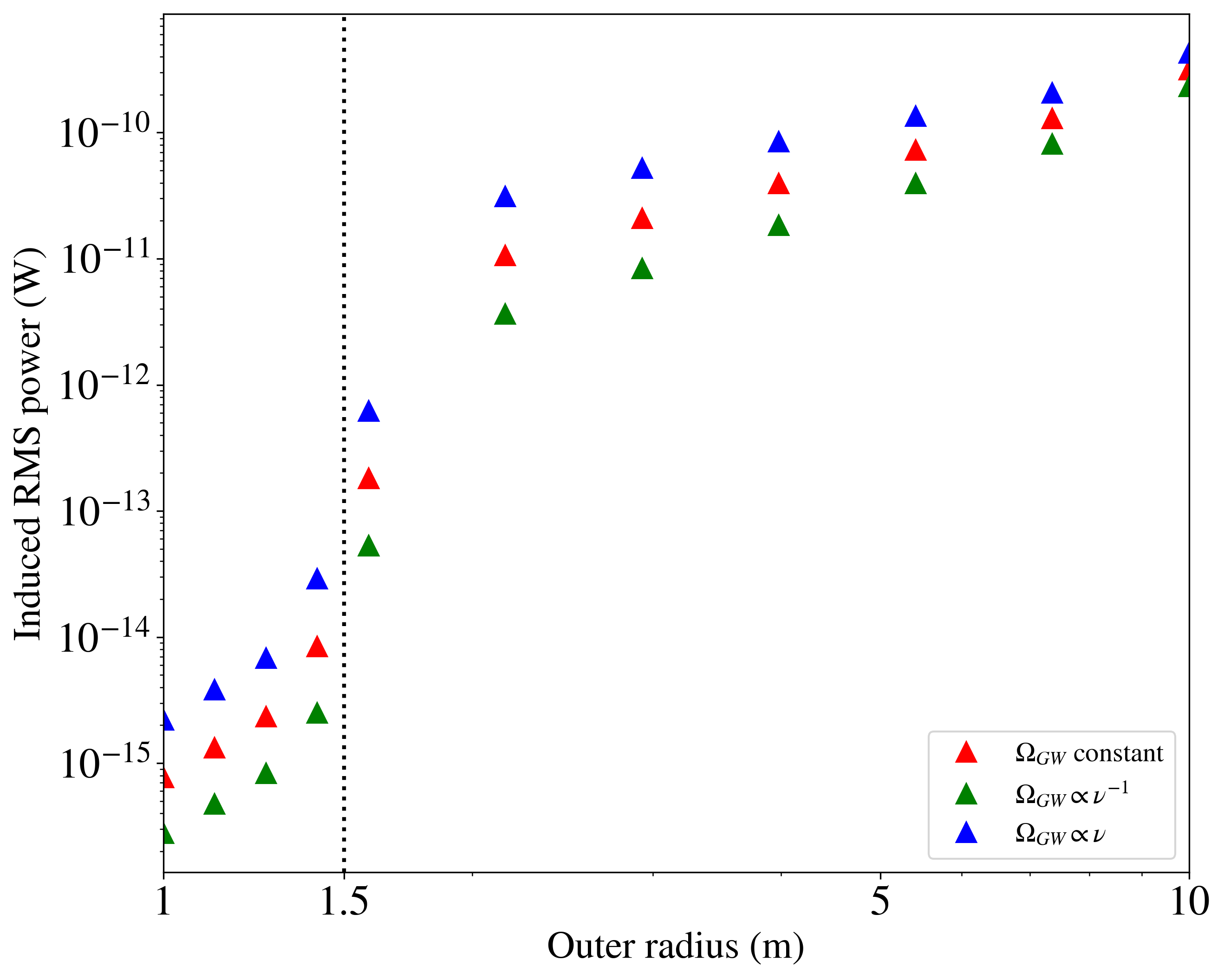}
    \caption{rms induced EM power generated in our cavity with respect to the radius of the detection scheme for our stochastic GW background toy model. We can see a jump in the response by two orders of magnitude due to the excitation of resonant frequencies. Below this point the resonant frequencies are above the cut-off frequency. After this point, the slope of the line is 2 for constant $\Om_{GW}$, which means that the factor that increase the response is the increasing of volume in the cavity. }
    \label{fig:radius}
\end{figure}

Considering the properties of hypothetical SGWB sources described earlier, we decided to consider a toy model that parameters could be adapted if we want to focus on some specific early Universe mechanism. The toy model for the power spectral density of our SGWB for any frequency $\nu$ is
\be
S_h(\nu)=\frac{3H_0^2}{4\pi^2}\Omega_{\rm GW}(\nu) \nu^{-3} sigm(-\nu+\nu_{\rm cut}),
\label{eq:toymodel}
\ee
where $H_0$ is the Hubble parameter of today, $\Omega_{\rm GW}(\nu)$ is the GW energy density per logarithmic frequency interval, $\nu_{\rm cut}$ is the cut-off frequency. At this cut-off frequency, the signal decreases exponentially as in~\cite{Maggiore2} due to the sigmoid function $sigm(\cdot)$.
Gathering the values of the different sources in~\cite{review}, we choose a constant $\Om_{GW}=10^{-10}$ and a cut-off frequency at 100 MHz for the toy model. This toy model is very useful to show how we could potentially use electromagnetic cavities to detect SGWB. However, real-world spectra for the SGWB should not be as simple as this toy model. The frequency disctribution of the spectrum could be not as smooth as in this toy model, or the cut-off could not so sharp. An example of a realistic GW spectrum is given by the SMASH model described in \cite{Ringwald:2022xif}, which combine the SGWB coming inflation, preheating and thermal fluctuations at the beginning of the radiation-dominated epoch. A first result from this toy model is that we can use an analogue of the Eq.~(\ref{eq:P(w)}) to compute the power spectral density (PSD) of the induced EM power in our cavity. The method is that we replace the temporal Fourier transform of the signal by the temporal Fourier transform of its autocorrelation.
The plot of this PSD can be found for three different cut-off frequencies in Fig.~\ref{fig:psd}. We can see that for our cavity, with the same parameters than in Fig.~\ref{fig:detlim}, the resonant frequencies before the cut-off will have much higher density than the other ones. This means that the auto-correlation of several signals coming from the stochastic background will be higher at the first resonant frequency and if we detect a significant drop between two resonant frequencies, the cut-off frequency must be somewhere in between. 
We insist on the fact that the possibility to tune the cavities parameters to get the resonant frequencies we want, for optimal detection. One cavity parameter that we can tune is the radius of the cavity, that change the resonant frequency. We show in Fig.~\ref{fig:psdr}. Modifying the radius of the cavity can help to spot the cut-off frequency when the cut-off frequency is between two resonant frequency. The cut-off frequency will be more difficult to spot otherwise. Moreover, using simultaneously cavities of different size could also help to spot multiple cut-off frequencies or even specific frequencies where the frequency dependence of GW spectrum changes.
This is also shown in Fig.~\ref{fig:radius}.
Here we simulate the rms induced power for several outer radius between 1m and 10m. We also considered other distribution for $\Om_{GW}$, as one can see through the green and blue points. There is a gap in the values around 1.5m. This gap is where the resonant frequencies goes below the cut-off frequency and the response is several order of magnitude higher. If one can combine cavities of different radii, we can also spot more precisely the cut-off frequency and also check whether or not  $\Omega_{GW}$ is constant with frequency. The value of the slope after the gap is 2 for constant $\Omega_{GW}$, that means that the main contribution of the increasing induced power with the radius comes from the increase of the detection volume, which is proportional to the square of the radius for a cylindrical cavity. Another assumption for $\Om_{GW}$ will lead to another slope in this model. For instance, for the $\Om_{GW}$ varying linearly with the frequency, the slope becomes equal to $1.5$. With these results one can also checking the frequency dependency of the GW spectrum, the presence of one or multiple cut-off frequencies or special frequencies where the frequency distribution change.
\\
\\
Please note that these results computed for the SGWB was made under one hidden assumption. In order to excite the resonant frequencies in the cavity, we shall assume that the SGWB has enough spatial and temporal coherence. By extension, the results presented above should be applied to any GW source that has enough coherence to trigger the resonance.
\\
\\
To sum up the results presented above, resonant detectors allow measuring the cut-off frequency of the yet hypothetical SGWB and also check the constancy of the GW density with frequency, two parameters that are specific to some early Universe SGWB.

\section{Conclusion and Discussion} In this work, we develop the frequency response analysis of resonant EM detectors of high-frequency GWs first described in~\cite{Herman2021} and references therein. The present work completes previous ones by giving tools for computing the response in function of time or frequency, depending on the application. While temporal approach is more suitable for transient signals like inspiralling binary primordial black hole mergers, the frequency approach is more suited to the search of the SGWB. In addition, this brand new frequency approach gives us an analytical, straight-forward and maybe more intuitive way to compute the response of the considered detection scheme, and fixed the mixed frame modeling mentioned in the literature. It can also examine directly the strain sensitivity at any given frequency after setting the experimental design and study the noise sources. This approach is also suitable theoretically for any GW signal. We just need the Fourier transform for inspiral signals, and strain power spectral density for stochastic ones. We also confirm that choosing the cavity parameters is very important to define the frequency range where we want an optimal detection. This location of the resonant frequencies is also very important to emphasize the cut-off frequency for the SGWB. The detection scheme response will get a massive drop above this special frequency. If the studied model has no sharp cut-off as described here, we could possibly spot a change in the frequency distributon at a specific frequency. The possibility to combine several cavities of different radii is also very interesting because you can get resonant response of different frequency bands. These kind of detectors are also a good addition to interferometers to detect SGWB at different frequencies. We also get a more realistic model since we take account of the losses at the walls of the cavity, therefore smoothing the resonance peaks.
%o there is no more divergence at the resonant frequencies.
%With this work, we complete the purely theoretical work on the cavity and the main perspective of this work is to take this detector with an experimental aspect. 
\\
\\
We claim the present work give strong arguments in favor for resonant EM detectors in the crucial SGWB investigation. Further studies should investigate deeper the experimental feasibility of first order induced power detection, as well as specific noise sources for this particular application. Another point of interest for further consideration is how one can correlate several detections to recreate the PSD of stochastic signals. 
By the way, we have provided some important prospects for motivating further experimental work.
%The fact that we use a macrosocopic first order response is a difference compared with some recent proposals. 
% je ne suis pas sur que je mettrais cela ici, car tu vas devoir citer l'autre article et entrer dans la polemique.
The results presented here indicate that
the expected strain sensitivity of such resonant detector could theoretically go below $10^{-30}$ for some frequencies. 
Moving toward an experiment based on the results presented here will require
 common effort from scientific communities working on axion detection, high-precision physics in strong magnetic field environment and high-frequency gravitational wave.
 We are confident resonant EM detectors of high-frequency gravitational waves will one day become the antennas through which we will listen to the most ancient relic of the Big Bang, glimpsing even further at the very origins of the Universe.\\\\
\begin{acknowledgments}
This research used resources of the "Plateforme Technologique de Calcul Intensif (PTCI)"
(\url{http://www.ptci.unamur.be}) located at the University of Namur, Belgium, which is supported
by the F.R.S.-FNRS under the convention No. 2.5020.11. The PTCI is member of the "Consortium des Équipements de Calcul Intensif
(CÉCI)" (\url{http://www.ceci-hpc.be}).
\end{acknowledgments}
\newpage
\appendix
\renewcommand{\theequation}{\Roman{equation}}

% reset the counter
\setcounter{equation}{0}
\section*{Appendix : Computing the RMS impulse response}
In this paper, we summarized how we can compute the induced electromagnetic field at first order, in the proper detector frame, in a gravitational wave detection process. This is the response we chose to compute for our detection schemes. This response can be function of time of frequency, but the duration of the signal or its frequency content could be quite different with respect of the incoming gravitational wave. This is why we have to compute a quantity that can be useful to compare signals between each other. For such oscillating field, we have to consider the root-mean-square average, that can be defined as
\begin{equation}
	\mathcal{P}_{\rm RMS}^2= \lim_{T\rightarrow\infty} \frac{1}{T} \int_{0}^{T} \left(\mathcal{P}(t) \right)^2 dt.
	\label{eq:prms}
\end{equation}

This is a key quantity because it is related to the $\mathcal{L}^2$ norm of the signal,  which is unitary for the Fourier transform. Physically, as it can be done for AC electric currents, the RMS power is related to the energy of the signal. One quantity that we can compute using this definition of RMS power is the RMS power impulse response. This the RMS power when the incoming signal is a sine wave. If we consider $h_+(t)= \sin(\omega t)$, combining the equation (\ref{eq:approx}) with equation~(~\ref{eq:sk10f}) and the inverse Fourier transform we have that 
\begin{equation}
	\hat{s}_{k10} (t)
	=\frac{7\pi B_0 I_k \omega}{3c} \sin\left(\frac{\omega L}{2c}\right)\sin\left(\omega t\right). 
	\label{eq:srir}
\end{equation}
This equation~(\ref{eq:srir}) is coherent with equation~(\ref{eq:sk10f}). With such a source term the solution of equation (\ref{eq:harmosc}) is simpler,
\begin{equation}
	\begin{aligned}
		\hat{b}_{k10} (t)=&e^{-\frac{\alpha_k c}{2Q} t} \left(A_{k10}  \cos\left(\varpi_k t\right) + B_{k10}  \sin\left(\varpi_k t\right)\right)\\ &+ \frac{7\pi c B_0 I_k \omega}{3} \sin\left(\frac{\omega L}{2c}\right) \frac{\sin(\omega t +\phi)}{\alpha_k^2c^2+\frac{i\omega\alpha_k c}{Q}-\omega^2}.
	\end{aligned}
\label{eq:solk10t}
\end{equation}
If we look at the behavior of this solution when $t$ goes to infinity, one can discard the homogeneous solution because of the decreasing exponential $e^{-\frac{\alpha_k c}{2Q} t}$. In this case, we can compute the $P_{\rm RIR}$, the RMS impulse response electromagnetic power, 
\begin{equation}
	P_{\rm RIR}(\omega)=\frac{7\sqrt{2}c}{3\mu_0}\pi^2 B_0^2 \omega^2\left\vert\sin\left(\frac{\omega L}{2c}\right)\right\vert \mathcal{S},
	\tag{\ref{eq:RIR}}
\end{equation}
where
\be
 \mathcal{S} =\sum_{k} \frac{\left( I_k\right)^2}{\sqrt{\left(\alpha_k^2c^2-\omega^2\right)^2+\left(\frac{\alpha_k \omega c}{2Q}\right)^2}}\cdot
 \tag{\ref{eq:S}}
\ee
This equation can be useful because if we neglect the contribution of the homogeneous solution that vanish when $t$ goes to infinity, equations~(\ref{eq:P(w)}) and~(\ref{eq:solk10t}) can show us that
\begin{equation}
	\vert P(\omega) \vert^2 = 2 \vert P_{\rm RIR}(\omega) \vert^2 \vert \tilde{h}_+(\omega) \vert^2,
	\label{eq:norm2}
\end{equation}
where we have here the GW power spectral density $\vert \tilde{h}_+(\omega) \vert^2$. This expression will help us to close this chapter by finding a practical estimation of RMS induced power when we have a signal $\tilde{h}_+(\omega)$. Let us first assume that $\mathcal{P}(t)$ is a periodic function, one can expand it as a real-valued Fourier series,
\begin{equation}
	\mathcal{P}(t) = \sum_{n=1}^\infty \left(a_n \cos\left(\tfrac{2\pi}{U} nt \right) + b_n \sin\left(\tfrac{2\pi}{U} nt \right) \right)
\end{equation}
where $U$ is the period. Putting the expression above in the definition~(\ref{eq:prms}) gives us
\begin{equation}
	\mathcal{P}_{\rm RMS}^2= \sum_{n=1}^\infty \frac{a_n^2+b_n^2}{2}.
\end{equation}
By generalizing the equation above to any function with the Fourier transform and using the approximation~(\ref{eq:norm2}), we find our estimation of the RMS power which is 
\begin{equation}
	\mathcal{P}_{\rm RMS}^2\simeq \int_0^\infty \vert P_{\rm RIR}(\omega) \vert^2 \vert \tilde{h}_+(\omega) \vert^2 d\omega.
	\tag{\ref{eq:RMSestim}}
\end{equation}
\onecolumngrid
\newpage
\section*{Appendix: Numerical validation of Eq.(\ref{eq:approx2})}
\begin{figure}[hb]
	\centering
	\includegraphics[scale=0.4]{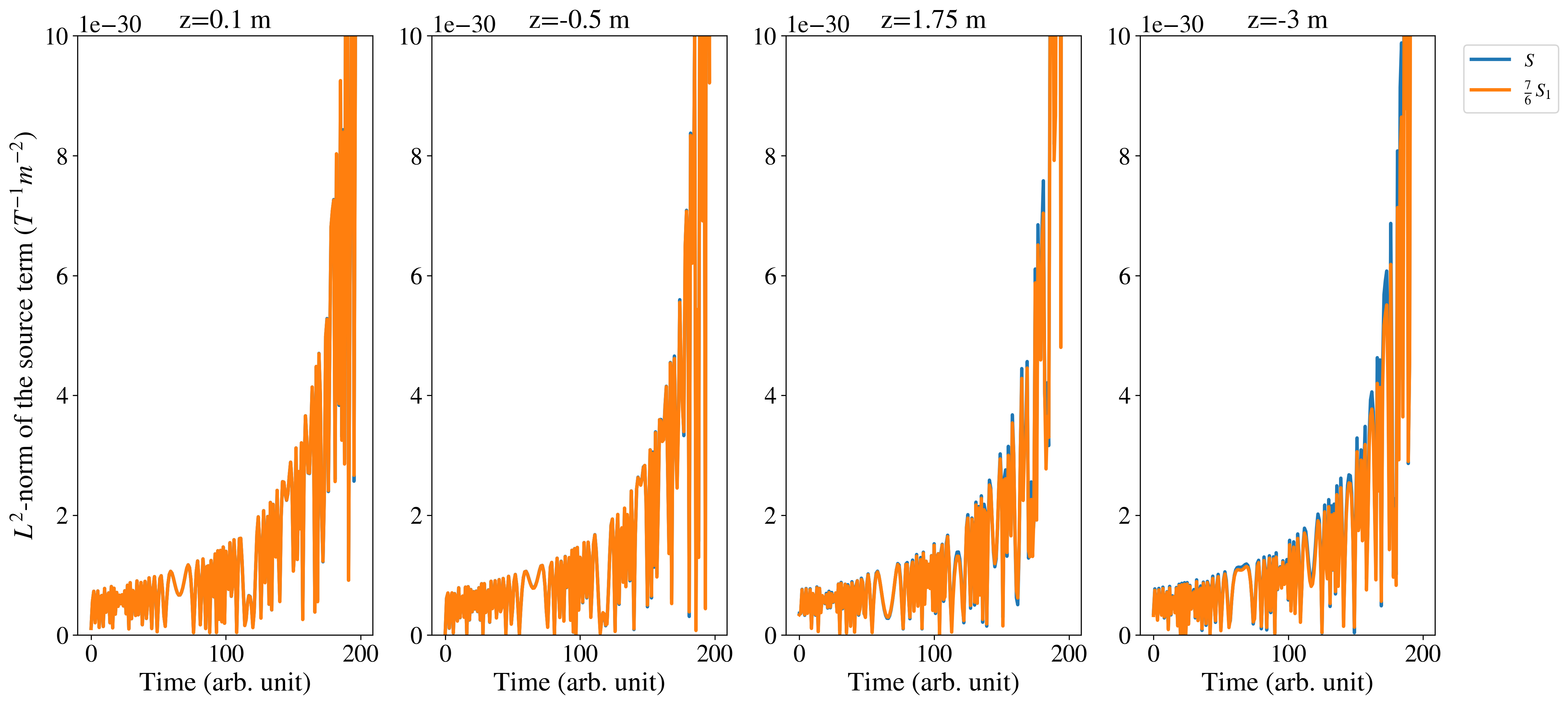}
	\includegraphics[scale=0.4]{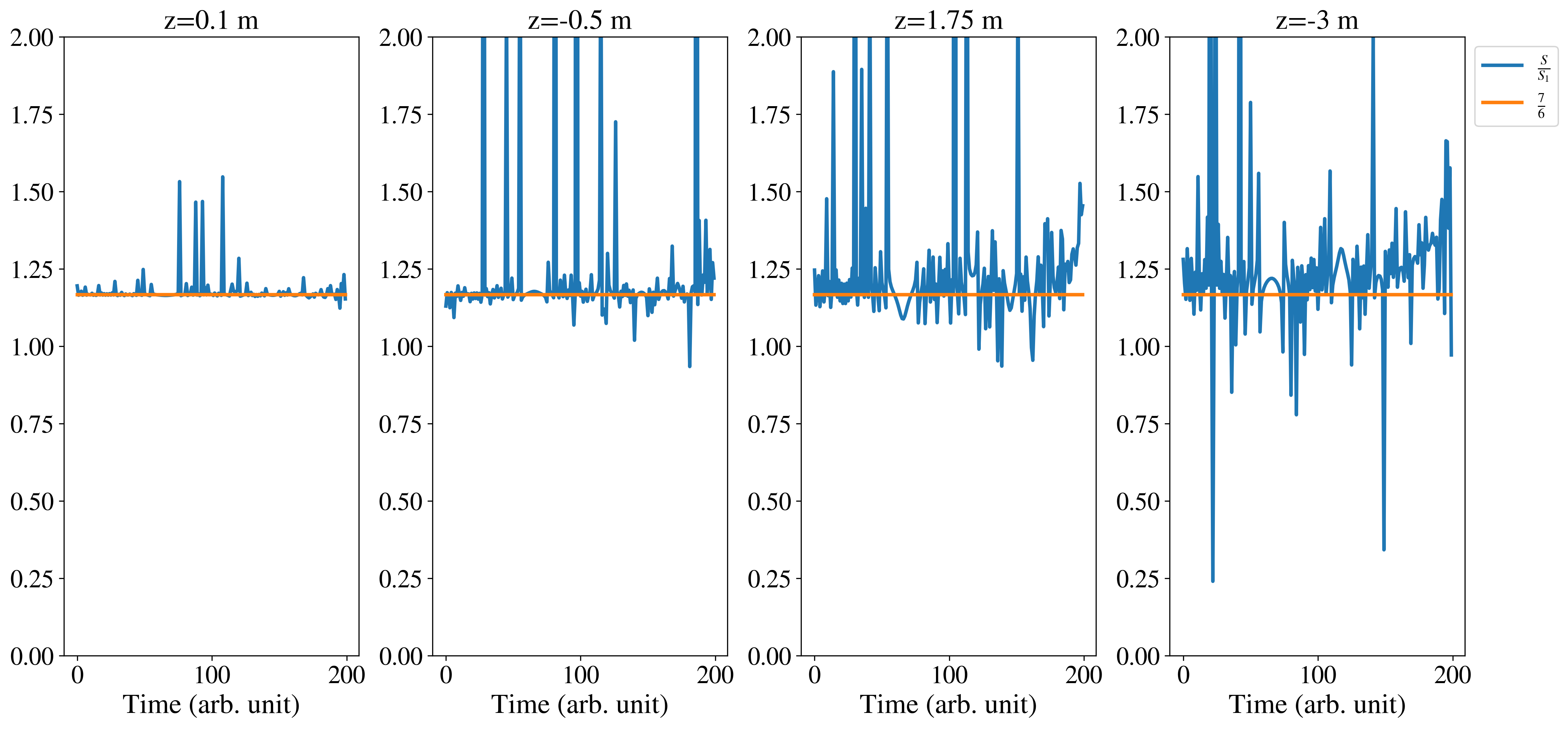}
	\caption{Numerical validation of the approximation Eq.~(\ref{eq:approx2}). On the top, the blue curve represents the $L^2$-norm of the source term for a slice at several values of $z$ (in meters), integrated over a unit disk for the x and y coordinates. The orange curve is the same computation for the approximation equation~(\ref{eq:approx}).The bottom panel is the source term in the detector frame divided by the source term in the TT gauge to validate equation~(\ref{eq:approx}).  The signal considered here is the Newtonian GW inspiral phase of primordial black hole mergers ($10^{-5} M_\odot$) you can find in \cite{Maggiore2007}. The number of terms considered in Eqs.~(\ref{eq:P},\ref{eq:Q}) are $n=10.$ Higher order terms will only modify the directional sensitivity of our detection scheme. Here we consider only the propagation along the symmetry axis of the cavity.}
	\label{fig:my_label}
\end{figure}
\twocolumngrid
\bibliographystyle{apsrev4-1}
\bibliography{biblio} 
\end{document}